\documentclass[12pt]{iopart}

\usepackage{cite}

\newcommand{\q}{{\bf q}}

\renewcommand{\l}{\delta l}

\begin{document}

\title[The $\varepsilon$-expansion in the symmetry-broken phase
of an interacting Bose gas]
{The $\varepsilon$-expansion in the symmetry-broken phase
of an interacting Bose gas at finite temperature}

\author{G Metikas, O Zobay and G Alber}

\address{Institut f\"ur Angewandte Physik, Technische Universit\"at Darmstadt,
64289 Darmstadt, Germany}

\begin{abstract}
We discuss the application of the momentum-shell
renormalization group method to the interacting homogeneous Bose gas in the symmetric and in the symmetry-broken
phases.
It is demonstrated that recently discussed discrepancies
are artifacts of not taking proper care of infrared divergencies appearing at finite temperature.
If these divergencies are taken into account and treated properly by means of the $\varepsilon$-expansion, the resulting renormalization group equations and the corresponding universal properties are identical in the symmetric
and the symmetry-broken phases.
\end{abstract}


\section{Introduction}

Renormalization techniques have been employed for the study of interacting Bose
gases near the critical temperature, because in this temperature regime
 the fluctuations dominate the mean field. The bulk of this work was written
 before the experimental realization of Bose-Einstein condensation (BEC) in ultracold atomic gases
\cite{Singh1,Singh2,Singh3,Singh4,Lee1,Lee2,CreswickWiegel,FisherHohenberg}.
It was shown that only at zero temperature the quantum nature of the
three-dimensional Bose gas differentiates it from a three-dimensional
two-component classical field theory. At any finite temperature the
Bose gas converges to the classical theory as the fixed point of
the renormalization group (RG) equations is approached.

Therefore the calculation of the universal properties of the Bose gas
 can be performed in the same way as for a classical theory, that is
 through the $\varepsilon$-expansion where $\varepsilon=4-D$ and $D$ is the
 number of spatial dimensions, e.g.
\cite{Wilsonfeynmann, WilsonFisher, WilsonKogut,WegnerHoughton}. In $D=3$,
although the results of the expansion up to second order in
$\varepsilon$ are in remarkable agreement with experimental values of
 critical exponents (measured in ${\rm He}^{4}$ experiments, but
 due to universality applicable in the case of Bose gases as well
\cite{Zinn}), higher-order results
 diverge from the experimental values. The reason
 is that the $\varepsilon$-expansion is asymptotic, as first noted in
 \cite{BrezinLeGuillou}, and
to obtain meaningful results when higher orders in $\varepsilon$
are included, one has to make use of resummation techniques, see
e.g. \cite{Kleinert}. This way critical exponents have been
 calculated up to fifth order in \cite{Vladimirov, Chetyrkin1,
 Chetyrkin2, Chetyrkin3, Chetyrkin4, Kazakov, Gorishny}, see also for
 corrections \cite{Kleinert1, Kleinert2} and improvements
 \cite{GuidaZinnJustin}. However, the results thus
 obtained are sensitive to the way the
resummation is performed and consequently somewhat ambiguous,
see e.g. \cite{LeGuillou}. 

An alternative method is to
calculate the universal properties perturbatively as series of powers
of  $g^{*}$ ($g^{*}$ being the
infrared stable fixed point for the interaction $g$) directly in
$D=3$, as first suggested in \cite{Parisi}. These
series are then truncated to order $g^{*L}$ where $L$ is the number of loops in which the
calculation is performed.
Though this method is fundamentally less
satisfactory than the $\varepsilon$-expansion, see
 e.g. \cite{Kleinert}, it can be used in the regime of small but
non-zero chemical potential. It has been employed for the calculation
of critical exponents up to seventh order in $g^{*}$ for $N=0,1,2,3$
\cite{LeGuillou2,BNM,LeGuillou} and for arbitrary $N$
\cite{Antonenko1,Antonenko2} where $N$ is the number of components of
the vector field. The series in $g^{*}$ are again asymptotic and have
to be resummed. There is in general agreement with the corresponding
$\varepsilon$-expansion results.
We will be referring to this technique as the direct method.

After the experimental
 realization of BEC in ultracold atomic gases \cite{AndEnsMat95,DavMewAnd95,BraSacTol95}, because of the renewed interest in these systems,
a new generation of papers on the renormalization of Bose gases appeared.
Starting with \cite{StoofBijlsmaRen }, a series of papers relied on the
so-called momentum-shell approach
\cite{Andersen, Alber, AlberMetikas, Metikas}.
In this method, momentum shells around the cutoff are successively
integrated out directly at $D=3$ according to Wilson's method, but unlike in the
direct method no expansion of the critical exponents over $g^{*}$ is
performed.

This apparently new method, when applied in the symmetric (normal)
 phase, yields universal results which, when compared to experimental values,
 are worse than even the first-order $\varepsilon$-expansion
 results. However, when the momentum-shell method is applied to the
 symmetry-broken phase, it yields results which are far better than the
 first-order $\varepsilon$-expansion and, in fact, as good as the results
 of the second-order $\varepsilon$-expansion. Based on this observation, it
 was assumed that the reliability of the momentum-shell method
 increases when it is used in the
 symmetry-broken phase, and a calculation of
 non-universal properties (for example transition temperature versus
 scattering length) from the symmetry-broken phase RG equations was
attempted.
For this reason, one may now wonder if applying the $\varepsilon$-expansion or the direct
method in the symmetry-broken rather than in the symmetric phase as is
usually done would
improve the results of these methods.

Here we show that the distinction introduced by the momentum-shell
method of
\cite{StoofBijlsmaRen, Andersen, Alber, AlberMetikas, Metikas}
 between RG methods applied in the symmetric and the symmetry-broken phases
 is an artifact of not taking care of the infrared divergence
which appears at finite temperature in the interacting Bose gas
theory. When this
divergence is taken into account and
treated properly, as in a classical theory, by means of the
$\varepsilon$-expansion, the RG equations and the resulting universal
properties are identical in the symmetric and in the symmetry-broken
phases.

Furthermore, even if one focuses on the regime of small but
non-zero chemical potential, where the direct method applies, the
use of the direct method in the symmetry-broken rather than
the symmetric phase
deteriorates the results instead of improving them.

\section{Derivation of RG equations}
 The partition function of the homogeneous $s$-wave interacting Bose gas is
\begin{equation}
 Z(\mu,\beta,V,g) \equiv {\rm Tr} e^{-\beta(\hat{H} - \mu \hat{N})} =
\int \delta[\phi,\phi^*] e^{-S[\phi,\phi^*]}
\label{partition}
\end{equation}
with the Euclidean action
\begin{equation}
\fl S[\phi, \phi^*] =  \frac{1}{\hbar}
 \int_0^{\hbar \beta } d\tau \int_{V} d^D {\bf x}
\left\{ \phi^* (\tau, {\bf x})
\left[\hbar\frac{\partial}{\partial \tau} - \frac{\hbar^2}{2m} \nabla^{2} - \mu\right]
 \phi(\tau, {\bf x}) + \frac{g}{2} |\phi(\tau, {\bf x})|^4
 \right\}.
\label{action}
 \end{equation}
We give an outline of the basic steps of the renormalization procedure. More
details can be found in, e.g., \cite{StoofBijlsmaRen,Alber,Metikas}.

In order to implement the first step of the RG procedure (Kadanoff
transformation), we split the field $\phi(x)$  into
a long-wavelength component $\phi_<(x)$  (slow field) and a short-wavelength
 component $\delta \phi_>(x)$ (fast field).
The fast field involves Fourier
 components which are
contained only in an infinitesimally thin shell in momentum space of
thickness $\Lambda e^{- \l} \leq | {\bf p}|\leq \Lambda$
near the momentum cutoff $\Lambda $, whereas
the slow field has all its Fourier components in
 the sphere whose center is at the origin of the momentum space and whose radius
 is $\Lambda e^{- \l }$.

We now perform the one-loop calculation of the effective theory of the
slow field. We integrate out the fast field and expand the resulting
effective action in powers of $g$ keeping up to order $g^2$.  This
perturbative effective action is equal to the original
action (\ref{action}) (with the
field $\phi $ replaced by the slow field $\phi_{<}$) plus two additional
terms. The first of these additional term is proportional to $g$ and
therefore quadratic in the modulus of the slow field. The second
additional term is proportional to $g^2$ and quartic to the modulus of
the slow field. The effective
Lagrangian of this theory can be cast in the
 form of the original Lagrangian because the additional terms produced by the
 integration over the fast fields have such a form that they can be considered
 as corrections to $\mu$ and $g$. Thus, after one
 infinitesimal integration, the chemical potential and the interaction in the
 effective action are
\begin{eqnarray}
\mu' & = \mu + d\mu =& \mu
-  g d_{D} \int_{\delta V_{{\bf p}}} dp~p^{D-1} f_{1}[\beta, E(p)-\mu],
  \nonumber \\
g' &= g + dg =& g - g^2 d_{D}
\int_{\delta V_{{\bf p}}} dp~p^{D-1}
f_{2}[\beta, E(p)-\mu]
\label{flows}
\end{eqnarray}
where $E(p) = p^2/2m$ and
\begin{eqnarray}
f_{1}(\beta, E) &=& 2 N(\beta E) +1, \nonumber \\
f_{2}(\beta, E) &=&
\frac{1+ 2 N(\beta E)}{ 2E} + 4
\beta N(\beta E) [1 + N(\beta E)]
\label{f1f2}
\end{eqnarray}
with the Bose-Einstein distribution $N(x)=1/(e^{x}-1)$.
The density of states $d_{D}$ can be expressed in terms of the
surface $\Omega_{D}$ of a unit hypersphere in $D$ dimensions according to
\begin{equation}
d_{D}=\frac{1}{(2\pi)^D} \Omega_{D} \hspace{0.5cm}
{\rm with} \hspace{0.5cm} \Omega_{D}=\frac{2 \pi^{D/2}}{\Gamma(D/2)}.
\end{equation}
The infinitesimal momentum shell which is integrated out is denoted by $\delta V_{{\bf p}}$.
We note that the integration procedure has no effect on the
inverse temperature or the slow field, i.e.,
\begin{equation}
\beta'=\beta, \ \phi_{<}'=\phi_{<}.
\end{equation}
Now the only remaining difference between the form of the effective action
and that of the original action is that in the original action the momentum of
the field is integrated from $0$ to the momentum cutoff $\Lambda$ whereas in
the effective action the momentum of the
slow field $|\q '|$ is integrated from $0$ to $\Lambda e^{- \l}$. This
difference is eliminated by a trivial rescaling of the momentum
$|q(\l )|= |\q '| e^{\l }$ which induces a trivial rescaling of the rest
of the parameters of the effective action,
 \begin{equation}
 \mu(\l)=\mu' e^{2\l},\ g(\l)=g' e^{ (2-D) \l},\ \beta(\l) = \beta e^{-2\l},
\ \phi_{<}(\l )= \phi_{<} e^{D \l /2}.
\label{tr}
\end{equation}
In terms of the rescaled parameters, equations (\ref{flows}) assume the form
\begin{eqnarray}
 \mu(\l)  & = & e^{2 \l } \mu
-  e^{2 \l } g  d_{D} \int_{\Lambda  e^{- \l } }^{\Lambda} dp~ p^{D-1}
f_{1}[\beta, E(p)-\mu],  \nonumber \\
g(\l)  & = & e^{(2-D) \l } g - e^{(2-D) \l } g^2  d_{D}
 \int_{ \Lambda e^{- \l } }^{\Lambda } dp~ p^{D-1} f_{2}[\beta, E(p)-\mu].
\label{rflows}
\end{eqnarray}
The system of (\ref{rflows}) becomes autonomous if one solves it
together with
\begin{equation}
\beta(\l)=e^{-2 \l} \beta.
\end{equation}
We perform the integrations in (\ref{rflows}) over the infinitesimal momentum shell $\delta V_{{\bf p}}$
 and keep terms only up to first order in $\l $, i.e.,
\begin{eqnarray}
 \mu(\l)  & = &  \mu + 2 \mu \l - g d_{D} \Lambda^{D} f_{1}[\beta, E(\Lambda) - \mu]
 \l,  \nonumber \\
g(\l)  & = &  g + (2-D) g \l - g^2 d_{D} \Lambda^{D} f_{2}[\beta,
E(\Lambda) - \mu] \l.
\end{eqnarray}
  Repeating the above procedure of
integrating out shells of
high-momentum and rescaling, we find the RG equations for the chemical
potential and the interaction,
\begin{eqnarray}
\frac{d \mu(l)}{dl}
 & = &  2 \mu(l)  - g(l) d_{D} \Lambda^{D} f_{1}[\beta(l),
E(\Lambda) - \mu(l)],
\nonumber \\
\frac{dg(l)}{dl}  & = &  (2-D) g(l)  - g(l)^2 d_{D} \Lambda^{D}
f_{2}[\beta(l), E(\Lambda) - \mu(l)]
\label{rgeq}
\end{eqnarray}
where the number of renormalization steps $l$ is a continuous parameter
running from $0$ (no shells integrated out) to $l^{*}=\infty$ (all shells
integrated out). The trivially rescaled quantities defined in
(\ref{tr}) also become functions of $l$.

Setting $g=0$ and consequently $g(l)=0$ in (\ref{rgeq}) one can study
 the ideal Bose gas both at zero ($\beta=\infty$) and at finite
 temperature ($\beta \neq \infty$), e.g. \cite{AlberMetikas}. One can also
 study the zero-temperature, interacting Bose gas, e.g.
\cite{Kolomeisky1, Kolomeisky2, FisherHohenberg, Sachdev}.  Finally, one can
 study the problem in its full generality, the
 finite-temperature interacting Bose gas.

In reference \cite{FisherHohenberg},
 the symmetric phase of the finite-temperature interacting Bose gas in
 an arbitrary number of dimensions was considered. It was shown that,
 for any finite temperature, the interacting Bose gas is driven
 towards a two-component classical system, as the fixed point of
 the RG equations is approached, see also
\cite{Singh1, Singh2, Singh3, Singh4, Lee1, Lee2, CreswickWiegel}.

Our formalism is different from that of
\cite{FisherHohenberg} or \cite{Singh1, Singh2, Singh3, Singh4, Lee1, Lee2, CreswickWiegel}, so it may be of some
interest to show that we can come to the same conclusion. In the
course of the following calculation we will also see how the finite
temperature theory we are examining here develops an infrared
divergence.

The RG method which we have been using is perturbative over the
interaction $g$. Therefore, as in the zero-temperature interacting gas
case, e.g. \cite{FisherStellenbosch}, we examine the behaviour of (\ref{rgeq}) near the
fixed point $(\mu^{*},\beta^{*})=(0,0)$ of the unperturbed system,
that is the finite-temperature
ideal gas. Near this fixed point the Bose-Einstein distribution can be
expanded as follows
\begin{equation}
N[\beta(E(p)-\mu)]=\frac{1}{\beta[E(p)-\mu]},
\end{equation}
and therefore equations (\ref{rflows}) near the fixed point become
\begin{eqnarray}
\fl \hspace*{0.4cm}  \mu(\l)  & = & e^{2 \l } \mu
-  e^{2 \l }  g
 d_{D} \int_{\Lambda  e^{- \l } }^{\Lambda} dp~ p^{D-1}~
 2 \frac{1}{\beta [E(p)-\mu]},    \nonumber \\
\fl \hspace*{0.4cm} g(\l)  & = & e^{(2-D) \l } g - e^{(2-D) \l } g^2  d_{D}
 \int_{ \Lambda e^{- \l } }^{\Lambda } dp~ p^{D-1}~
\left\{\frac{9}{2} \frac{1}{E(p)-\mu} + 5 \frac{1}{\beta (E(p)-\mu)^2}
  \right\}.
\end{eqnarray}
We observe that, at $(\mu=0,\beta=0)$, the last term
 in each of the above equations is infrared
 divergent for all $p$ and all $D$. This divergence can be treated by
redefining variables; we recast the above equations in
terms of the new variable ${\tilde g}=g/\beta$,
\begin{eqnarray}
 \mu(\l)  & = & e^{2 \l } \mu
-  e^{2 \l } {\tilde g}\beta   d_{D}
\int_{\Lambda  e^{- \l } }^{\Lambda} dp~p^{D-1}~
f_{1}[\beta, E(p)-\mu],  \nonumber \\
{\tilde g}(\l)  & = & e^{(4-D) \l } {\tilde g} - e^{(4-D) \l } {\tilde
g}^2 \beta  d_{D}
 \int_{ \Lambda e^{- \l } }^{\Lambda } dp~p^{D-1}~f_{2}[\beta, E(p) - \mu]
\label{tflows}
\end{eqnarray}
which can be also written in a differential form
\begin{eqnarray}
\frac{d \mu(l)}{dl}
 & = &  2 \mu(l)  - {\tilde g}(l) \beta(l)
d_{D} \Lambda^{D}  f_{1}[\beta(l), E(\Lambda)-\mu(l)],
\nonumber \\
\frac{d{\tilde g}(l)}{dl}  & = & (4-D) {\tilde g}(l)  -
{\tilde g}(l)^2 \beta(l) d_{D} \Lambda^{D} f_{2}[\beta(l), E(\Lambda)-\mu(l)].
\label{dclrgeq}
\end{eqnarray}
These are exactly the equations (4.6) of \cite{FisherHohenberg}
for the chemical potential and the interaction supplemented by equations (4.12)
 of \cite{FisherHohenberg} in the classical regime.

Near the unperturbed fixed point $(\mu^{*},\beta^{*})=(0,0)$,
Eqs.\ (\ref{tflows}) reduce to
\begin{eqnarray}
 \mu(\l)  & = & e^{2 \l } \mu
-  e^{2 \l }  {\tilde g}
 d_{D} \int_{\Lambda  e^{- \l } }^{\Lambda} dp~ p^{D-1}~
  2 \frac{1}{ E(p)- \mu }, \   \nonumber \\
{\tilde g}(\l)  & = & e^{(4-D) \l } {\tilde g} - e^{(4-D) \l } {\tilde
g}^2  d_{D}
 \int_{ \Lambda e^{- \l } }^{\Lambda } dp~ p^{D-1}~ 5 \frac{1}{ [E(p)-\mu]^2}.
\label{exclrgeq}
\end{eqnarray}
We note that apart from some numerical coefficients, equations (\ref{exclrgeq})
are identical with the equations ensuing from the
classical Landau-Ginzburg-Wilson (LGW) reduced Lagrangian, see,
e.g., equations (6.40)-(6.42) of \cite{FisherStellenbosch}.

The different coefficients are due to the fact that the Bose gas
theory involves a complex field, whereas the classical LGW theory uses
a two-dimensional vector field. As we will see later, this difference
is insignificant in the sense that the two theories have the same
universal behaviour.

Finally, it is convenient to cast (\ref{dclrgeq})
 in terms of dimensionless variables
$M=\beta_{\Lambda} \mu$,
 ${\bar G}=m^2 \Lambda^{D-4} {\tilde g} d_{D} / \hbar^4$,
 $E_{>}=\beta_{\Lambda} E(\Lambda)= 1/2$
 (so that we keep track of the energy terms), and
$b=\beta/\beta_{\Lambda}$,
with $\beta_{\Lambda}=m/(\hbar^2 \Lambda^{2})$, which yields
\begin{eqnarray}
\frac{d M(l)}{dl}
 & = &  2 M(l) - {\bar G}(l) b(l) f_{1}[b(l), E_{>} - M(l)],
\nonumber \\
\frac{d{\bar G}(l)}{dl}  & = & (4-D) {\bar G}(l)  -
{\bar G}(l)^2 b(l) f_{2}[b(l), E_{>} - M(l)].
\label{clrgeq}
\end{eqnarray}
It may not be immediately obvious that the variable
 ${\bar G}$ is dimensionless. However recalling that
  $g=\Omega_{D} \hbar^2 a^{D-2}/m$, see \cite{Wiegel}, we can rewrite
 it as ${\bar G} = \Omega_{D} (a \Lambda)^{D-2} /b$ which is
 clearly dimensionless.

\subsection{$\varepsilon$-Expansion}

Since equations (\ref{exclrgeq}) have the same structure as the classical RG
 equations, the same difficulties in the infrared regime appear. In
 particular around $\mu=0$ and for $D<4$ the integral in the equation for the
interaction is infrared divergent. It is interesting to note that this
 divergence originates in the terms
\[\frac{N[\beta (E(p)-\mu)]}{E(p)-\mu} \hspace{1cm} {\rm and}
 \hspace{1cm} N[\beta (E(p)-\mu)]^2 \] of $f_{2}[\beta, E(p) - \mu]$,
see equation (\ref{f1f2}). As in the classical case, this
divergence is cured by performing the $\varepsilon$-expansion.
We identify $4-D$ in the equation for the interaction with
 $\varepsilon$. Furthermore we assume that $\mu$ and ${\tilde g}$ (and
 therefore $M$ and ${\bar G}$) are of
 the same order and expand the RG equations (\ref{clrgeq}) up to
 second order in these variables which yields
\begin{eqnarray}
\fl \hspace*{0.4cm} \frac{d M(l)}{dl}
 & = &  2 M(l) - {\bar G}(l) b(l)  \left\{ 2  N[b(l) E_{>}] + 2
 b(l) N[b(l) E_{>}] (1+N[ b(l) E_{>} ]) M(l) \right\},
\nonumber \\
\fl \hspace*{0.4cm} \frac{d{\bar G}(l)}{dl}  & = & \varepsilon {\bar G}(l)  -
{\bar G}(l)^2 b(l)
\left\{ \frac{1+2N[b(l) E_{>} ]}{2E_{>}} + 4 b(l)
N[b(l) E_{>}] (1+N[b(l) E_{>}]) \right\}.
\label{epsrgeq}
\end{eqnarray}
This system has a trivial fixed point at
$(M^{*},{\bar G}^{*})=(0,0)$ with eigenvalues
\begin{eqnarray}
\lambda_{1}&=&2, \nonumber \\
\lambda_{2}&=&\varepsilon = 4-D.
\end{eqnarray}
Therefore, the eigenspace of $\lambda_2$ corresponds to the
unstable direction for $D<4$, to the marginal one for $D=4$, and to the stable
direction for $D>4$. There is also a non-trivial fixed point
\[  (M^{*}, {\bar G}^{*}) = \left[ \frac{ \varepsilon }{ 10 - 2 \varepsilon},
\frac{5 \varepsilon }{(10 + 2 \varepsilon)^2 } \right] = \left[ \frac{\varepsilon}{10}+O(\varepsilon^2),
\frac{ \varepsilon }{20} + O(\varepsilon^2) \right]. \]
Up to first order in $\varepsilon$, its eigenvalues are
\begin{eqnarray}
\lambda_{1}&=&2-\frac{2}{5}\varepsilon, \nonumber \\
\lambda_{2}&=& - \varepsilon = D-4,
\label{epseigen}
\end{eqnarray}
and consequently the eigenspace of $\lambda_2$ corresponds to the stable direction for $D<4$,
to the marginal one for $D=4$, and to the unstable direction for $D>4$.

 For the case of physical interest, $D=3$, we set $\varepsilon=1$ in the
 expressions which we have already expanded up to first order in
 $\varepsilon$. A simple example of a universal property that we can now
 calculate is  the critical exponent for the correlation length $\nu=1/\lambda_{1}=0.600 + O(\varepsilon^2)$.
This is exactly the same
 as the result found in the $\varepsilon$-expansion study of a classical
 two-component LGW theory in $D=3$,
e.g. \cite{Wilsonfeynmann, WilsonFisher, WilsonKogut},
see also \cite{CreswickWiegel}. A finite temperature interacting Bose
 gas in three  spatial dimensions belongs to the same universality class
 as a two-component classical field theory in three spatial
 dimensions \cite{CreswickWiegel}.

\subsection{Direct Method}
We set $D=3$ directly in (\ref{rgeq}) or equivalently in
(\ref{clrgeq}). The fixed point is
$$(\mu^{*}, {\tilde g}^{*})=
(E(\Lambda)/6, 5\pi^2 E(\Lambda)^2/(18\Lambda^3))$$
 or
equivalently in dimensionless units $(M^{*},{\bar
G}^{*})=(1/12,5/144)$.
 We then linearize around the fixed point and
calculate the eigenvalues
\begin{eqnarray}
\lambda_{1}&=& \frac{1}{50}~ \left(-1728 {\bar G}^{*} + \sqrt{2985984
{\bar G}^{*2} +
57600 {\bar G}^{*} + 625 } + 75 \right), \nonumber \\
\lambda_{2}&=& \frac{1}{50}~ \left(-1728 {\bar G}^{*} - \sqrt{2985984
{\bar G}^{*2} +
57600 {\bar G}^{*} + 625 } + 75 \right).
\end{eqnarray}
Finally, we expand the critical exponent $\nu=1/\lambda_{1}$ up to first order in ${\bar G}^{*}$
because we are performing a one-loop calculation,
\[ \nu=\frac{1}{2} + \frac{72 {\bar G}^{*}}{25} + O({\bar G}^{*2}) = 0.600 +
O({\bar G}^{*2}). \]
This is the same result as for a two-component classical theory, see
e.g. \cite{Antonenko1, Antonenko2}.

\subsection{Momentum-Shell Method}
This approach is used in \cite{StoofBijlsmaRen }, see also
 \cite{Alber, Andersen}, and corresponds to
 setting $D=3$ directly in (\ref{rgeq}) or equivalently in
 (\ref{clrgeq}) as in the direct method. In this case the
 RG equations (\ref{rgeq}) coincide with equations (11a) and (11b)
 of \cite{StoofBijlsmaRen }. The fixed point is the same as in the
 direct method. However now we set ${\bar G}^{*}=5/144$ in the eigenvalues
\begin{eqnarray}
\lambda_{1}&=& \frac{3 + \sqrt{249}}{10}, \nonumber \\
\lambda_{2}&=& \frac{3 - \sqrt{249}}{10}
\end{eqnarray}
and without expanding we find
 $ \nu = 1/\lambda_{1} \approx 0.532$.

\section{Symmetry-broken phase}

If we spontaneously break the global ${\rm U}(1)$ symmetry of
(\ref{action}) by introducing the most probable configuration
 $\overline{\phi}$, we find
\begin{eqnarray}
\fl \hspace*{1cm} S[\phi, \phi^{*}] =&&
-\beta V \left[\mu n_{0} - \frac{n_{0}^2 g}{2}\right] \nonumber \\
\fl  && + \frac{1}{\hbar}
 \int_0^{\hbar \beta } d\tau \int_{V} d^D {\bf x} ~\phi^{*}(\tau, {\bf x})
 \left[ \hbar~\frac{\partial}{\partial \tau} - \frac{\hbar^{2}}{2m}~\nabla^{2}
 -\mu + 2 g n_{0}\right] \phi (\tau, {\bf x}) \nonumber \\
&& + \frac{g n_{0}}{2\hbar}~\int_0^{\hbar \beta } d\tau \int_{V} d^D
{\bf x}
 ~\left[\phi^{*}(\tau, {\bf x})
\phi^*(\tau, {\bf x}) +
\phi(\tau, {\bf x}) \phi(\tau, {\bf x}) \right] \nonumber \\
&& + \frac{g \overline{\phi}}{\hbar}~ \int_0^{\hbar \beta } d\tau \int_{V} d^D {\bf x} ~\left[ \phi^{*}(\tau,
{\bf x})
\phi^{*}(\tau, {\bf x}) \phi(\tau, {\bf x}) + \phi^{*}(\tau, {\bf x})
 \phi(\tau, {\bf x}) \phi(\tau, {\bf x}) \right] \nonumber \\
&& + \frac{g}{2\hbar}~\int_0^{\hbar \beta } d\tau \int_{V} d^D {\bf x}
  ~\phi^{*}(\tau, {\bf x}) \phi^{*}(\tau,
{\bf x}) \phi(\tau, {\bf x}) \phi(\tau, {\bf x})
\label{brokenaction}
\end{eqnarray}
where
$n_{0}=|\overline{\phi}|^{2}=\mu/g$ is the condensate density, and
$\phi (\tau, {\bf x})$ now denotes the fluctuation around the most
probable configuration $\overline{\phi}$.

It is possible to write down the RG equations for the symmetry-broken
phase. The calculation is significantly more complicated than in the symmetric
phase, for details on the derivation of the RG equations
see \cite{StoofBijlsmaRen} for $D=3$
and \cite{Alber, AlberMetikas} for arbitrary $D$. We focus here on the
RG equations for the chemical potential and the interaction
\begin{eqnarray}
\fl \hspace*{1.5cm} \frac{d M(l)}{dl}
 & = &  2 M(l) - {\bar G}(l) b(l)  \left\{  \frac{2 E_{>}^3+6 M(l)
 E_{>}^2 + M(l)^3}{2 \Delta(l)^3} [2 N(b(l) \Delta(l) ) + 1] - 1
  \right. \nonumber \\
\fl &  & +\left.  \frac{M(l)(2 E_{>} + M(l))^2 }{\Delta(l)^2 } b(l)
N(b(l) \Delta(l)) [N(b(l)
 \Delta(l)) + 1] \right\},
\nonumber \\
\fl \hspace*{1.5cm} \frac{d{\bar G}(l)}{dl}  & = & (4-D) {\bar G}(l) -
{\bar G}(l)^2 b(l)
\left\{ \frac{ ( E_{>}- M(l) )^2 }{2 \Delta(l)^3 } [2N(b(l)
 \Delta(l))+1] \right. \nonumber  \\
&& +\left.
 \frac{(2E_{>}+M(l))^2}{\Delta(l)^2} b(l) N(b(l) \Delta(l))[N(b(l)
\Delta(l))+1]
 \right\}
\label{sbrgeq}
\end{eqnarray}
where $\Delta(l) =\sqrt{E_{>}^2 + 2 M(l) E_{>}}$.
 We note that the above RG equations
coincide with (\ref{clrgeq}) of the symmetric phase for $M=0$.

\subsection{$\varepsilon$-Expansion}

At the fixed point $(\mu^{*}, \beta^{*}) = (0,0)$
of the unperturbed system an infrared divergent term appears in the
course of the derivation of the RG equation for the interaction. This
term is exactly the same as in the symmetric case, i.e., \[
\int_{ \Lambda e^{- \l } }^{\Lambda } dp~ p^{D-1}~ 5 \frac{1}
{ (E(p)-\mu)^2},\] but now originates in the terms
\[  \frac{(E(p)-\mu)^2}{\delta(p) ^3} N(\beta \delta(p) )
\hspace{1cm}  {\rm and}  \hspace{1cm} \frac{(2 E(p) + \mu)^2 }{\delta(p)
^2}  N(\beta \delta(p))^2   \] with $\delta(p) =
\sqrt{E(p)^2 + 2 \mu E(p) }. $

We can easily apply the $\varepsilon$-expansion technique to equations
(\ref{sbrgeq}) in order to cure the infrared divergence.
As we have seen in the symmetric phase, all we have to do is
identify $4-D$ with $\varepsilon$ in the equation for the
interaction and then expand the chemical
potential equation up to first order in $M$ and the equation for the
interaction up to zeroth order in $M$. We thus find the $\varepsilon$-expansion
RG equations
\begin{eqnarray}
\fl \hspace*{1.2cm}  \frac{d M(l)}{dl}
  &=&   2 M(l) - {\bar G}(l) b(l)  \left\{ 2  N[b(l) E_{>}] + 2
 b(l) N[b(l) E_{>}] (1+N[ b(l) E_{>} ]) M(l) \right\},
\nonumber \\
\fl \hspace*{1.2cm}  \frac{d{\bar G}(l)}{dl}   &=& \varepsilon {\bar G}(l)  -
{\bar G}(l)^2 b(l)
\left\{  \frac{1+2 N[b(l) E_{>}]}{2 E_{>}} + 4 b(l)
N[b(l) E_{>}] (1+N[b(l) E_{>}]) \right\}. \nonumber \\
\label{sbepsrgeq}
\end{eqnarray}
Comparing (\ref{sbepsrgeq}) to (\ref{epsrgeq}) we see that they are
exactly the same! In other words, when we perform the momentum-shell
integrations together with the $\varepsilon$-expansion,
 the symmetric and symmetry-broken phases yield
exactly the same RG equations, and consequently identical
universal properties.

\subsection{Direct Method}
As in the symmetric phase, it consists of setting directly $D=3$ in
(\ref{sbrgeq}). The non-trivial fixed point is $(M^{*},{\bar G}^{*})=
(1/2,1/4)$. The
eigenvalues of (\ref{sbrgeq}), when it is linearized around the fixed
point, are
\begin{eqnarray}
\lambda_{1}& = & \frac{9 \pi^2- 12 {\bar G}^{*} + \sqrt{3} (-16 {\bar
G}^{*2} + 24 \pi^2 {\bar G}^{*} +3 \pi^4 )^{1/2} }{6 \pi^2}, \nonumber \\
\lambda_{2}& = &  \frac{9 \pi^2- 12 {\bar G}^{*} - \sqrt{3} (-16 {\bar
G}^{*2} + 24 \pi^2 {\bar G}^{*} +3 \pi^4 )^{1/2} }{6 \pi^2},
\end{eqnarray}
and therefore \[ \nu = 0.500 + O({\bar G}^{*2}). \]

\subsection{Momentum-Shell Method}

We set  directly $D=3$ in
(\ref{sbrgeq}) as in the direct method. The resulting RG equations,
when recast in the dimensionful variables $\mu$ and $g$,
coincide with (29a), (29b) of \cite{StoofBijlsmaRen}, see also
equations (16), (17) of \cite{Alber}. We now
substitute ${\bar G}^{*}$ in the eigenvalues of the direct method
\begin{eqnarray}
\lambda_{1}&=&\frac{3+\sqrt{33}}{6}, \nonumber \\
\lambda_{2}&=&\frac{3-\sqrt{33}}{6},
\end{eqnarray}
and without expanding we find
$\nu=1/\lambda_{1}\approx 0.686$.

\section{Comparison and Conclusion}

The first-order $\varepsilon$-expansion and the momentum-shell
method used in \cite{StoofBijlsmaRen} and in subsequent papers
\cite{Andersen,Alber,AlberMetikas} do not yield identical universal
properties. As we have seen, the first-order $\varepsilon$-expansion gives
the value $\nu=0.600 + O(\varepsilon^2)$ for the critical exponent of the
correlation length both in the symmetric and in the symmetry-broken
phases. However the momentum-shell method gives $\nu=0.532$ in the
symmetric and $\nu=0.686$ in the symmetry-broken phase. These results
are to be compared to the experimental value $\nu=0.670$, see
e.g. \cite{Zinn}. It seems as if the momentum-shell method in the
symmetry-broken phase yields the best result (in fact, almost as good
as the second order in the $\varepsilon$-expansion result $\nu=0.655 +
O(\varepsilon^3)$, see e.g. \cite{Kleinert}).
 However, the results do not always justify the method used to obtain them.
For example, in subsection 2.1, had we not
first expanded $\nu=1/\lambda_{1}$ up to first order in $\varepsilon$ and
then set $\varepsilon=1$ but used the unexpanded formula (\ref{epseigen}) for
$\lambda_{1}$, we would have found
$\nu=0.625$. Although this is closer than $\nu=0.532$
to the experimental result $\nu=0.670$, it is clearly incorrect.

The main point of this paper is that the momentum-shell method results
depend on whether the RG calculation is performed in the symmetric or the
symmetry-broken phase, whereas the first order $\varepsilon$-expansion
results do not. The dependence of the momentum-shell results on the
phase is an artifact of not avoiding the infrared divergence of the
Bose gas theory. Because for any finite temperature the Bose gas
theory has the same infrared behaviour as a classical two-component
theory, the $\varepsilon$-expansion as we know it from classical papers,
e.g. \cite{WilsonKogut}, can cure the infrared divergence and
yield reliable results.

Furthermore we note that, even if we do not worry about the infrared
divergence, and use the direct method, applying it in the
symmetry-broken phase makes the result for the critical exponent of
the correlation length worse rather than improving it as one would
have expected from the momentum-shell method.

The above discussion and comparison of values does not take into
account the effect of several approximations (derivative expansion,
polynomial truncation of the effective action, assumptions about the
sharpness or smoothness of the infrared cutoff separating the fast
from the slow modes) which were employed in the course of the
calculation. For discussions on these approximations see \cite{Morristheexact,
Andersen, Metikas}. However since these approximations were used in
the part of the calculation which is common for all three methods,
that is the derivation of the RG equations (\ref{clrgeq}) (or equivalently
(\ref{rgeq}) ) and (\ref{sbrgeq}), a
comparison between these methods is still valid.

\ack
This work is supported by the Deutsche Forschungsgemeinschaft within the For\-scher\-gruppe `Quantengase'. Stimulating discussions with
G.\ Shlyapnikov are gratefully acknowledged.


\section*{References}
\begin{thebibliography}{10}

\bibitem{Singh1}
Singh K K 1975 \PL A {\bf 51}  27

\bibitem{Singh2}
Singh K K 1975 \PR B {\bf 12}  2819

\bibitem{Singh3}
Singh K K 1976 \PL A {\bf 57}  309

\bibitem{Singh4}
Singh K K 1976 \PR B {\bf 13}  3192

\bibitem{Lee1}
Lee J C 1979 \PR B {\bf 20}  1277

\bibitem{Lee2}
Lee J C 1980 {\it Physica} A {\bf 104}  189

\bibitem{CreswickWiegel}
Creswick R J and Wiegel F W 1983 \PR A {\bf 28}  1579

\bibitem{FisherHohenberg}
Fisher D S and Hohenberg P C 1988 \PR B {\bf 37}  4936

\bibitem{Wilsonfeynmann}
Wilson K G 1972 \PRL {\bf 28}  548

\bibitem{WilsonFisher}
Wilson K G and Fisher M E 1972 \PRL {\bf 28}  240

\bibitem{WilsonKogut}
Wilson K G and Kogut J 1974 {\it Phys.\ Rep.\ } {\bf 12}  75

\bibitem{WegnerHoughton}
Wegner F A and Houghton A 1973 \PR A {\bf 8}  401

\bibitem{Zinn}
Zinn-Justin J 1989, {\em Quantum Field Theory and Critical Phenomena} (Oxford: Oxford University Press)

\bibitem{BrezinLeGuillou}
Br\'ezin E, Guillou J C L, Zinn-Justin J and Nickel B G 1973 \PL
 A {\bf 44}  227

\bibitem{Kleinert}
Kleinert H and Schulte-Frohlinde V 2001, {\em Critical Properties of
  $\phi^{4}$-Theories} (Singapore: World Scientific)

\bibitem{Vladimirov}
Vladimirov A A, Kazakov D I and Tarasov O V 1979 {\it Sov.\ Phys.\ JETP} {\bf 50}
  521

\bibitem{Chetyrkin1}
Chetyrkin K G, Kataev A L and Tkachov F V 1981 \PL B {\bf 99}  147


\bibitem{Chetyrkin2}
Chetyrkin K G, Kataev A L and Tkachov F V 1981 \PL B {\bf 101}  457


\bibitem{Chetyrkin3}
Chetyrkin K G and Tkachov F V 1981 {\it Nucl.\ Phys.\ } B {\bf 192}  159

\bibitem{Chetyrkin4}
Chetyrkin K G, Gorishny S G, Larin S A and Tkachov F V 1983 \PL
 B {\bf 132}  351

\bibitem{Kazakov}
Kazakov D I 1983 \PL B {\bf 133}  406

\bibitem{Gorishny}
Gorishny S G, Larin S A and Tkachov F V 1984 \PL A {\bf 101}  120

\bibitem{Kleinert1}
Kleinert H, Neu J, Schulte-Frohlinde V, Chetyrkin K G and Larin S A
1991 \PL B {\bf 272}  39

\bibitem{Kleinert2}
Kleinert H, Neu J, Schulte-Frohlinde V, Chetyrkin K G and Larin S A
1993 \PL B {\bf 319}  545

\bibitem{GuidaZinnJustin}
Guida R and Zinn-Justin J 1998 \JPA {\bf 31}  8103

\bibitem{LeGuillou}
Guillou J C L and Zinn-Justin J 1980 \PR B {\bf 21}  3976

\bibitem{Parisi}
Parisi G 1980 {\it J.\ Stat.\ Phys.\ } {\bf 23}  49

\bibitem{LeGuillou2}
Guillou J C L and Zinn-Justin J 1977 \PRL {\bf 39}  95

\bibitem{BNM}
Baker G A Jr, Nickel B G and Meiron D I 1978 \PR B {\bf 17}  1365

\bibitem{Antonenko1}
Antonenko S A and Sokolov A I 1995 \PR E {\bf 51}  1894

\bibitem{Antonenko2}
Antonenko S A and Sokolov A I 1998 {\it Sov.\ Phys.\ Sol.\ State} {\bf 40}  1169

\bibitem{AndEnsMat95} Anderson M H, Ensher J R, Matthews M R, Wieman C and Cornell E A 1995 {\it Science} {\bf 269} 198

\bibitem{DavMewAnd95} Davis K B, Mewes M O, Andrews M R, van Druten N J, Durfee D S, Kurn D M and Ketterle W 1995 \PRL {\bf 75} 3969

\bibitem{BraSacTol95} Bradley C C, Sackett C A, Tollett J J and Hulet R G 1995 \PRL {\bf 75} 1687

\bibitem{StoofBijlsmaRen}
Bijlsma M and Stoof H T C 1996 \PR A {\bf 54}  5085

\bibitem{Andersen}
Andersen J O and Strickland M 1999 \PR A {\bf 60}  1442

\bibitem{Alber}
Alber G 2001 \PR A {\bf 63}  023613

\bibitem{AlberMetikas}
Alber G and Metikas G 2001 {\it Appl.\ Phys.\ } B {\bf 73}  773

\bibitem{Metikas}
Metikas G and Alber G 2002 \jpb {\bf 35}  4223

\bibitem{Kolomeisky1}
Kolomeisky E B and Straley J P 1992 \PR B {\bf 46}  11749

\bibitem{Kolomeisky2}
Kolomeisky E B and Straley J P 1992 \PR B {\bf 46}  13942

\bibitem{Sachdev}
Sachdev S 2001 {\em Quantum Phase Transitions} (Cambridge: Cambridge University Press)

\bibitem{FisherStellenbosch}
Fisher M E 1982 {\em Scaling, Universality and Renormalization Group Theory }
  (Lectures presented at the ``Advanced Course on Critical Phenomena" held at The
  Merensky Institute of Physics, University of Stellenbosch, South Africa)

\bibitem{Wiegel}
Wiegel F W 1975 {\it Phys.\ Rep.\ } {\bf 16}  57

\bibitem{Morristheexact}
Morris T R 1994 {\it Int.\ J.\ Mod.\ Phys.\ } A {\bf 9}  2411

\end{thebibliography}
\end{document}